\documentclass[]{article}
\usepackage[total={6.5in, 10in}]{geometry}
\usepackage[utf8]{inputenc}
\usepackage[super,sort,comma]{natbib}
\usepackage{authblk}
\usepackage{graphicx}
\usepackage{xcolor}
\usepackage{ulem}
\usepackage{url}
\usepackage{gensymb}
\usepackage{multirow} 
\usepackage{amsmath}
 \usepackage{booktabs}
\usepackage{textgreek}
\usepackage{lineno}

\title{Retrieval of multiple fibre orientations using X-ray dark-field signal modelling}

\author[1,a,*]{Lorenzo Massimi}
\author[1]{Michela Fratini}
\author[2]{Shashidhara Marathe}
\author[2]{Christoph Rau}
\author[3,4]{Giuseppe Gigli}
\author[5]{Alessandro Olivo}
\author[1]{Alessia Cedola}
\affil[1]{CNR-Nanotec (Institute of Nanotechnology), Rome, Italy}
\affil[2]{Diamond Light Source, Harwell Oxford Campus, OX11 0DE Didcot, UK}
\affil[3]{University of Salento, Lecce, Italy}
\affil[4]{Tecnomed Puglia - Tecnopolo per la medicina di precisione (Biotech Lecce Hub) c/o Campus Ecotekne, Lecce, Italy}
\affil[5]{Department of Medical Physics and Biomedical Engineering, University College London, UK}

\affil[a]{Current address: Fondazione Santa Lucia IRCCS, Rome, Italy}
\affil[*]{l.massimiphd@gmail.com}

\begin{document}

\maketitle

\begin{abstract}
Dark-field imaging is widely used to infer fibre orientation from signal modulation as a function of sample orientation. However, current X-ray dark-field retrieval methods are restricted to single orientations and require tomography to resolve overlapping structures. This approach is time-consuming and not suitable for thin materials, which are common in materials science. Here we present a dark-field model capable of retrieving multiple fibre orientations within a single pixel. The model, based on a geometrical description of fibre scattering, was validated through Monte Carlo simulations and experiments using beam-tracking setups with 1D and 2D masks. Results demonstrate reliable orientation retrieval for up to two fibres per pixel, with the 2D mask providing multi-directional sensitivity in a single acquisition and enabling faster and simplified data collection.
\end{abstract}

\section{Introduction}
The retrieval of fibre orientation is important in many fields, from medical imaging to materials science. In medical imaging, it is used to map white matter pathways, or to study muscle fibres, whose alterations are linked to several pathological conditions \cite{schmahmann2009fiber,douaud2011dti}. Fibre structure is also critical in carbon composites, widely used in the aerospace and automotive industries, where failure can have catastrophic consequences \cite{beaumont1989failure,heslehurst2014defects}. A single fibre direction is often assumed for simplicity, but in reality fibres overlap. This happens in brain tissue, and in composites, where layers are usually stacked at different orientations to improve mechanical properties.
Imaging fibres is challenging because it requires both high spatial resolution and a large field of view, which are rarely achieved simultaneously. X-ray microtomography can resolve fibres, but the required micron or sub-micron resolution restricts the field of view, providing only a limited view of the sample \cite{martin2007microstructure}. To overcome this limitation, indirect methods have been developed. Instead of directly visualizing fibres, these methods infer orientation from the modulation of a signal caused by fibre anisotropy. In medical imaging, diffusion tensor imaging (DTI) uses directional water diffusion to retrieve fibre orientation. A simple tensor model works for single orientations, while q-ball imaging extends the analysis to multiple orientations within a voxel \cite{hasan2011review,tuch2004q}. However, DTI cannot be applied to dry materials and is therefore unsuitable for materials science. 
X-ray based approaches offer broader applicability. Small Angle X-ray Scattering (SAXS) uses the higher scattering intensity perpendicular to the fibre direction to achieve orientation sensitivity \cite{georgiadis2021nanostructure,georgiadis2020retrieving}. However, the requirement of raster scans makes its application very slow when implemented with laboratory sources. Other laboratory-based X-ray methods such as grating interferometry, speckle imaging, edge-illumination, and beam tracking provide access to ultra-small-angle scattering (dark-field, DF). With these techniques, fibre orientation can be estimated from the modulation of the DF signal, but current analyses are limited to single orientations and require tomography to resolve overlaps \cite{jensen2010directional,malecki2014x,kagias20162d,smith2022x}. Tomography complicates the experimental setup, requires multiple degrees of freedom, and is poorly suited for thin samples like composite sheets or plates \cite{sharma2016six}. The complex acquisition scheme also restricts time-resolved studies, which are crucial for investigating damage dynamics. In this work, we present a DF signal model that enables the retrieval of multiple fibres layer orientations within a single pixel of a DF radiograph obtained using a beam-tracking (BT) setup. The model is tested through simulations and validated with experiments using two BT configurations: one with a 1D mask and one with a 2D mask, the latter greatly simplifying both the setup and the data acquisition.

\section{Methods}

\subsection*{Model}
In BT, the DF signal is described as the convolution of the incident beam with a scattering function $\Sigma$, which is approximated as a Gaussian since its main effect is a broadening of the beam \cite{endrizzi2014retrieval,jensen2010directional}. Therefore, for a 2D sample the scattering function can be written as
$\Sigma = \big(\begin{smallmatrix} 
\sigma_1^2 & 0\\
  0 & \sigma_2^2
\end{smallmatrix}\big)$ where the off-diagonal terms describing correlations are neglected. Here $\sigma_{1,2}$ can be interpreted as scattering power along and perpendicular to the fibre direction. Considering a single scattering layer with orientation $\varphi$ and a BT setup using a 1D mask for simplicity, the DF intensity for each pixel $(x,y)$ can be written as:
\begin{equation}
    I_{DF}(x,y,\theta) = \hat{v}\Sigma_\varphi\hat{v}^T
    \label{eq1}
\end{equation}
where $\Sigma_\varphi = R_\varphi\Sigma R_\varphi^T$, with $R_\varphi$ the rotation matrix for the angle $\varphi$, and $\hat{v} = (1,0)$ the direction of system sensitivity (perpendicular to the mask aperture). 

For multiple layers with the same scattering properties but orientations $\varphi_i$, the total scattering function can be considered as the sum of the scattering functions, as DF scales linearly with thickness \cite{doherty2023edge}. In addition, to retrieve the DF modulation, measurements at different sample angles $\theta_j$ are needed. A directional DF dataset with overlapping layers can therefore be written as:
\begin{equation}
    I_{DF}(x,y,\theta_j) = \hat{v}R_{\theta_j}\left(\sum_{i=1}^N\Sigma_{\varphi_i}\right)R_{\theta_j}^T\hat{v}^T
    \label{eq2}
\end{equation}
where $\theta_j$ varies in the range 0-180\textdegree . The use of 1D mask allows to probe one $\theta_j$ direction only, requiring sample rotation, while a 2D mask can probe several directions simultaneously, similarly to 1D and 2D gratings in grating-based imaging \cite{kagias20162d}.  

For $N=1,2,3$ layers, explicit expressions can be obtained by expanding Eq.\ref{eq2}:
\begin{align}
I_1(\theta) &= \tfrac{1}{2}(\sigma_1^2 + \sigma_2^2) 
+ \tfrac{\sigma_1^2 - \sigma_2^2}{2}\cos(2(\theta - \varphi_1)) \notag \\[8pt]
I_2(\theta) &= (\sigma_1^2 + \sigma_2^2) 
+ \tfrac{\sigma_1^2 - \sigma_2^2}{2} 
\Big[ \cos\big(2(\theta - \varphi_1)\big) + \cos\big(2(\theta - \varphi_2)\big) \Big] \notag \\[8pt]
I_3(\theta) &= \tfrac{3}{2}(\sigma_1^2 + \sigma_2^2) 
+ \tfrac{\sigma_1^2 - \sigma_2^2}{2} 
\Big[ \cos\big(2(\theta - \varphi_1)\big) + \cos\big(2(\theta - \varphi_2)\big) 
+ \cos\big(2(\theta - \varphi_3)\big) \Big]
\label{eq3}
\end{align}
The resulting DF intensity consists of a constant term that increases with the number of layers, and of a modulated term. The latter depends on the difference in the scattering function $\Sigma$ along the two principal directions (similar to fractional anisotropy in DTI \cite{figley2022potential}) and on the superposition of multiple oscillatory contributions. Once the number of layers per pixel is determined, fitting the measured DF intensity modulation with the proposed equations enables the retrieval of fibres orientation. Curve fitting was performed by fixing the values of $\sigma_{1,2}$, estimated from a single-layer ROI, with a tolerance of 10\%. Allowing $\sigma_{1,2}$ to vary freely did not yield a reliable retrieval. In addition, pixels corresponding to a single layer were fitted first, as their orientations are expected to be the most reliable. The fitting was then extended to the other regions, using the orientations of adjacent pixels as the initial guess.  

\subsection*{Simulated data}
Dark-field images of a carbon composite phantom were generated with McXTrace \cite{bergback2013mcxtrace}. The phantom consisted of three differently oriented layers of vertically extended graphite cylinders simulating monodirectional overlapped carbon fibres, arranged in a $300$ $\times$ $200$ $\mu$m$^2$ region, with diameters between 1 and 10 $\mu$m. The layers were rotated relative to each other by $\varphi_1=40^\circ$, $\varphi_2=90^\circ$, and $\varphi_3=130^\circ$. A parallel beam geometry was assumed, placing the sample a few meters away from the source with a monochromatic photon energy of 20 keV. A standard BT setup using a 1D thin perfectly absorbing mask was considered according to previous approaches \cite{millard2014monte}. Phase retrieval was performed by Gaussian fitting of the beamlet profiles according to conventional BT phase retrieval \cite{endrizzi2014retrieval}.

\subsection*{Experimental data}
\begin{figure*}[bht]
\includegraphics[scale = 1.11]{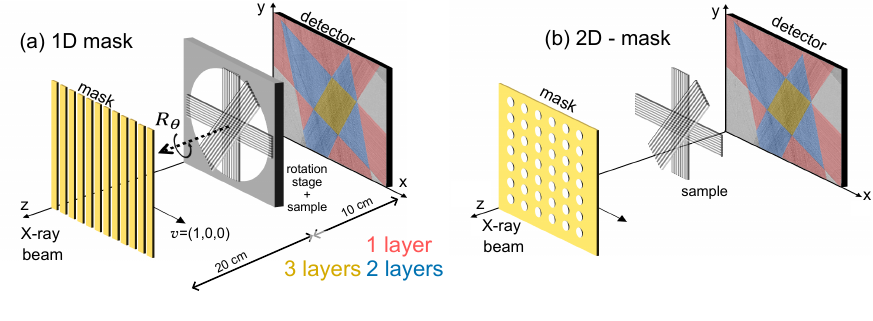}
\caption{(a, b) Schematic view of BT setup using 1D and 2D absorption mask, respectively.}
\label{fig0a}
\end{figure*}

Experimental X-ray DF data were acquired at I13 (imaging branch) beamline of the Diamond Light Source (UK), A pink-beam spectrum was employed in order to maximise the flux, with the energy peak around 20 keV \cite{rau2011coherent}. Two different experimental BT setups were implemented at the beamline and schematically illustrated in Fig.\ref{fig0a}.

The first setup used a 1D absorption mask with 20~$\mu$m apertures and 80~$\mu$m pitch. In the second setup, a 2D mask with circular holes of 20~$\mu$m diameter and 80~$\mu$m period was used providing multi-directions sensitivity. In both cases, aperture-limited spatial resolution was achieved with dithered acquisition, consisting of multiple acquisitions by translating the sample in sub-period steps \cite{diemoz2014spatial}. In the 1D case, 4 dithering steps perpendicular to the mask direction have been acquired, while the use of the 2D mask required four steps in each direction for a total of 16 steps. Conventional phase retrieval based on Gaussian profile curve fitting was applied to all images acquired with both the setups to separate the three contrast channels i.e. transmission, refraction, and dark-field\cite{endrizzi2014retrieval,vittoria2015beam}. In the 2D case, DF retrieval was performed by extracting multiple angular profiles from each hole and fitting the so obtained Gaussian curves, as in the previous case. Since high-frequency signal was observed in the dark-field images related to residual free-space propagation signal in the transmission channel\cite{massimi2021optimization}, a low-pass Gaussian blurring was applied before analysis. The detection was performed using an indirect conversion system based on a PCO edge 5.5 sCMOS camera coupled with a scintillator and microscopy optics. The combination resulted in an effective pixel size of 2.6 $ \mu$m. The sample consisted of three partially overlapping monodirectional carbon-fibre sheets, each with a thickness of about 0.2~$\mu$m. It is important to note that, due to the sample design, identical fibre orientations are expected across all regions within a given layer. This serves as a reliable reference for evaluating model performance in areas containing different numbers of layers.

\section{Results}
To test the validity and reliability of the proposed model, Montecarlo simulations were performed.
\begin{figure*}[bht]
\includegraphics[scale = 1.35]{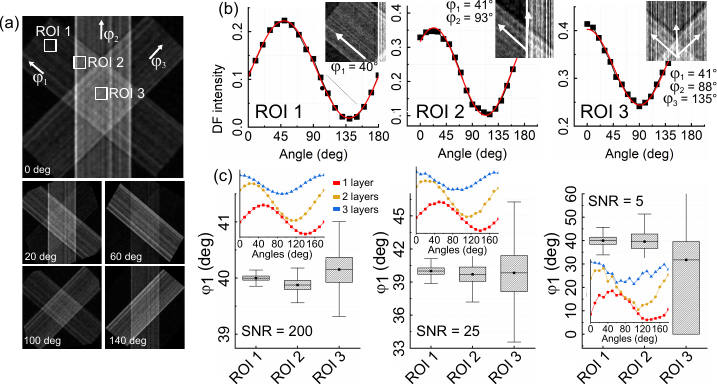}
\caption{(a) Simulated DF images of a carbon-composite phantom at different rotation angles. (b) Angular intensity profiles from the ROIs in (a) with retrieved fibre orientations overlapped to a crop of simulated image (inset). (c) Retrieved orientations for one, two, and three fibre ROIs at varying noise levels shown as box plots, with fitted profiles in the inset for signal quality comparison.}
\label{Fig0}
\end{figure*}
The resulting DF images at different rotation angles are shown in Fig.~\ref{Fig0}(a), clearly displaying signal modulation, with fibre visibility depending on their orientation relative to the mask apertures. As expected, the DF intensity scales linearly with the number of fibre layers, with three-layer regions producing stronger signals than single or double layer ones. The proposed DF model was applied to simulated ROIs containing one, two, and three layers. Results are shown in Fig.~\ref{Fig0}(b), where fits (red lines) are compared with data retrieved from the simulated images (black squares). For a single layer (ROI 1), the retrieved orientation matched the expected value. With two layers (ROI 2), the model still reproduced the correct directions but small deviations appeared, reaching about 5\textdegree for three layers (ROI 3). Insets show cropped images with retrieved orientations (white arrows) overlaid for visual comparison. To further evaluate stability, a Monte Carlo-based analysis was performed on repeated curve fitting. $N=1000$ intensity profiles were generated according to Eq.~\ref{eq3}, with Gaussian noise added before curve fitting. Note that performing this analysis on profiles extracted from simulated images would have required computing a large number of images (N images for each orientation angle), which would have taken a significant amount of time. Therefore, it has been preferred to directly generate the signal modulation curve. However, simulating realistic DF noise is challenging due to the complex error propagation that occurs during phase retrieval\cite{massimi2021optimization}. For simplicity, noise was added using a signal-to-noise ratio (SNR) defined as $\mu/\sigma$, where $\mu$ is the mean DF signal and $\sigma$ the standard deviation of the background. Given the curve average value, the definition above was used to estimate the gaussian standard deviation required to achieve specific SNR values. Results for direction $\varphi_1$ are shown in Fig.~\ref{Fig0}(c) and summarised in Table~\ref{tab0}. Example of profiles at the different SNR values are included as insets for visual comparison.  
\begin{table}[h]
\centering
\label{tab0}
\caption{Mean angles with standard deviations for different SNR levels across the three ROIs}
\begin{tabular}{cccc}
\toprule
 & \multicolumn{3}{c}{\textbf{Orientation Angle}} \\
\cmidrule(lr){2-4}
\textbf{} & ROI 1 & ROI 2 & ROI 3 \\
\midrule
200 & $40.0 \pm 0.1^\circ$ & $39.9 \pm 0.1^\circ$ & $40.1 \pm 0.3^\circ$ \\
25  & $40.0 \pm 0.4^\circ$ & $39.7 \pm 1.0^\circ$ & $38.2 \pm 8.3^\circ$ \\
5   & $39.9 \pm 2.5^\circ$ & $39.7 \pm 4.9^\circ$ & $29.3 \pm 15.6^\circ$ \\
\bottomrule
\end{tabular}
\label{tab0}
\end{table}

Plots and table show that single layer orientations can be retrieved with high accuracy, with errors below 2.5\textdegree even at low SNR. For multiple layers, the error increases with the  fibre number. This is expected, as fitting multiple overlapping oscillatory functions introduces variability since a phase shift in one curve can be partially compensated by others. At SNR = 25, directions from two layers regions are still retrieved reliably (error about 1\textdegree), but three layers regions start showing deviations of several degrees. At SNR = 5, the error rises to about 15\textdegree, making retrieval unreliable in the three layers case but still acceptable for one and two layers.

\begin{figure*}[htbp]
\includegraphics[scale = 1]{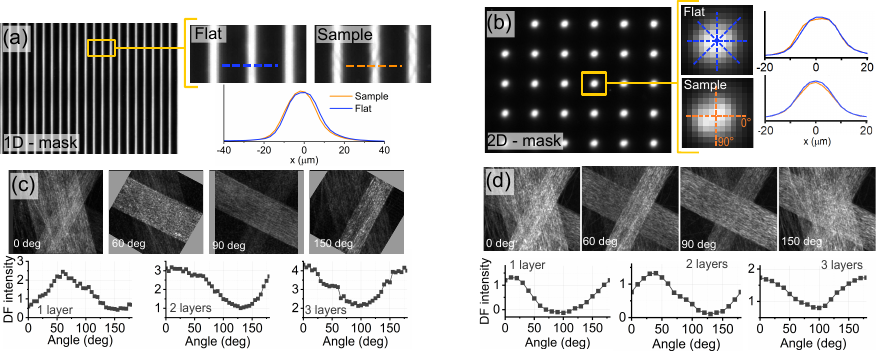}
\caption{Phase-retrieval procedures for both masks (a,b), and retrieved DF images at different mask–sample angles (c,d) with corresponding intensity profiles from ROIs containing one, two, or three layers.  
}
\label{Fig1}
\end{figure*}

To provide experimental validation of the proposed model, a directional BT experiment was carried out. The experimental images are shown in Fig.\ref{Fig1}.
Mask images with and without the sample for the 1D mask case, together with beamlets line profiles are reported in Fig.\ref{Fig1}(a) showing a change in the beamlets' profile as expected from the sample design. Retrieved DF images (Fig.~\ref{Fig1}(c)) show strong modulation due to the high scattering power of carbon fibres, with angular line profiles revealing phase shifts and average signal variations between regions containing one, two, or three layers. Intensity modulation is also observed for refraction, but not for transmission (see supplemental material Fig. S1 for images of the additional channels). As predicted by the proposed model, it can be observed that the average signal increases with the number of layers. The images for the BT setup employing a 2D mask are shown in Fig.~\ref{Fig1}(b). Circular apertures provide directional sensitivity in all the orientations simultaneously for the three contrast channels (see supplemental Material Fig.S1 for images of the additional channels), eliminating the need for sample rotation \cite{navarrete2023x} and greatly simplifying the experimental setup but at the cost of a longer acquisition time due to dithering acquisition in both vertical and horizontal directions. However, it is worth noting that this limitation can be removed by exploiting alternative 2D mask designs \cite{lioliou2022cycloidal}.  The retrieved DF images (Fig.~\ref{Fig1}(d)) show similar modulated intensity profiles to the linear case, as expected, but with reduced noise. This improvement arises because all angular profiles originated from the same dataset while in the linear case separate acquisitions of each angular step introduced uncorrelated fluctuations between measures.

\begin{figure*}[htb]
\includegraphics[scale = 0.87]{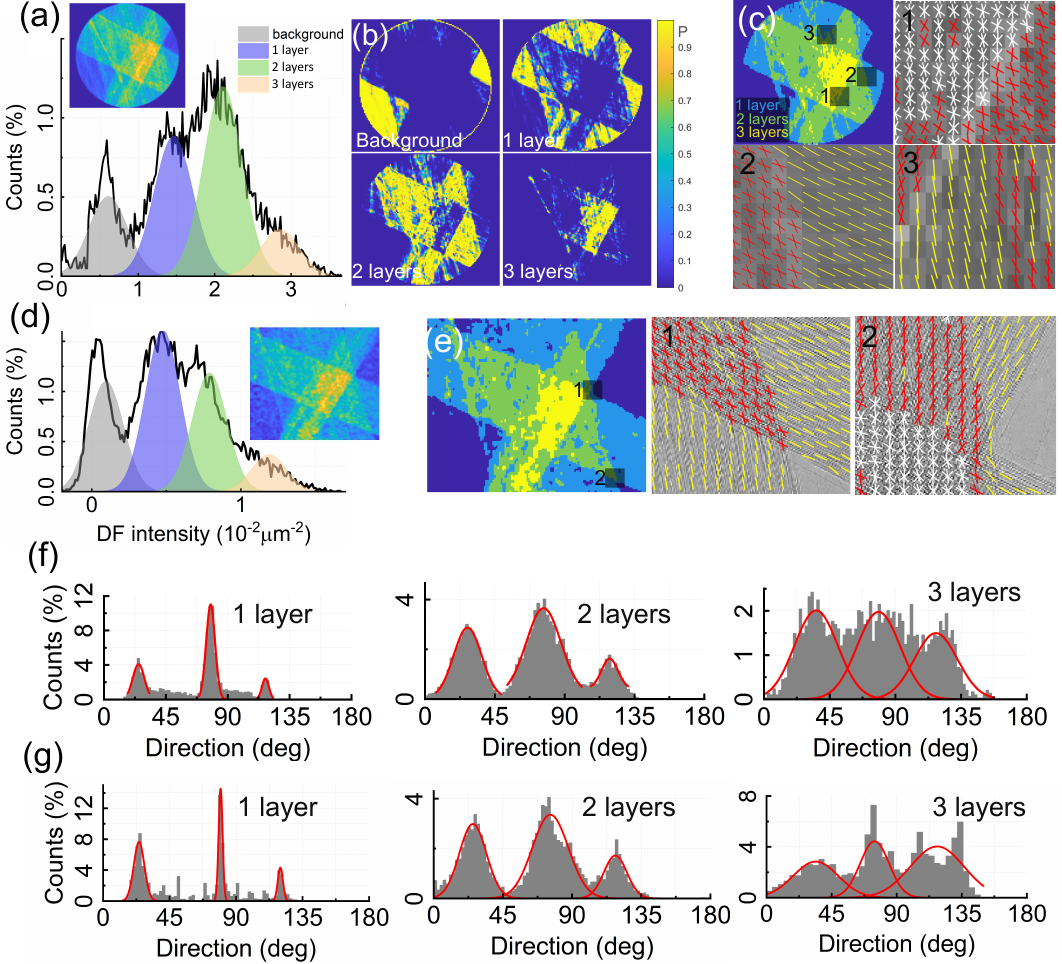}
\caption{Panels (a,d) show the fit of the mean dark-field intensity (insets) for the 1D and 2D masks, respectively. (b) probability maps for each pixel to belong to background, one, two, or three layer regions. (c,e) binarized maps assigning the most probable layer count to each pixel; retrieved fibre orientations for ROIs 1–3 are shown in the side panels (single fibre in yellow, two fibres in red, three fibres in white). (d,g) Quantitative comparison of fibre orientations retrieved from one, two, and three layer regions for the 1D and 2D masks, respectively.}
\label{Fig2}
\end{figure*}

The proposed directional model was applied to experimental data of the phantom acquired with the two setups described above, with results reported in Fig.\ref{Fig2}.  
The first challenge was to determine the number of layers contributing to each pixel, in order to select the appropriate fitting equation. As shown in Eq.~\ref{eq3}, the isotropic scattering component increases with the number of layers. Therefore, the average DF signal over all angles was used to estimate the layer count per pixel. In Fig.~\ref{Fig2}(a), the average DF map is shown together with the corresponding histogram of intensity values. The histogram was fitted with four gaussians corresponding to the background and three associated with regions containing one, two or three layers. This analysis provided, for each pixel, the probability of containing a given number of layers, as illustrated for the 1D mask case in Fig.~\ref{Fig2}(b). Similar maps are obtained for the 2D mask case with the corresponding histogram of the intensity values reported in Fig.\ref{Fig2}(d). A threshold (P $>$ 0.6) was then applied to assign the most likely number of layers to each pixel. The resulting layer-count maps are reported in Fig.~\ref{Fig2}(c,e) for the 1D and 2D masks, respectively, showing good agreement with the expected layer distribution from the sample design (supplemental Material Fig.S2). A complementary estimation based on fit-error analysis produced similar results (supplemental Material Fig.S3). Importantly, a close correspondence between data acquired independently with the two masks is observed. To retrieve fibre orientation, the proposed model was subsequently fitted pixel by pixel according to the assigned layer number. 
The resulting orientations are presented in Fig.~\ref{Fig2}(c,e) for the 1D and 2D masks, as magnified views of selected regions highlighted by numbered dark squares in the layer-count maps. Notably, the orientations remain consistent across transitions between regions with different layer counts, particularly evident in the ROIs highlighted in Fig.~\ref{Fig2}(e). 

A quantitative assessment of the retrieved orientations is provided in the histograms of fibre directions (Fig.~\ref{Fig2}(f),(g)) for the three different regions, with results summarized in Table~\ref{tab1}.  
\begin{table}[h]
  \centering
  \caption{Layers orientations retrieved in different regions for both experimental setups. Values are reported as mean $\pm$ standard deviation, obtained by fitting the distributions in Fig.~\ref{Fig2}(d,g).}
  \label{tab1}
  \setlength{\tabcolsep}{8pt}      
  \renewcommand{\arraystretch}{0.95} 
  \small                           
\begin{tabular}{l c c c }
    \toprule
    & $\varphi_1$\textdegree & $\varphi_2$\textdegree & $\varphi_3$\textdegree \\
    \midrule
    \addlinespace[0.8ex]
    \multicolumn{4}{c}{\textbf{1D mask}} \\
    \addlinespace[0.8ex]
    1 layer & $25.1\pm4.2$  & $77.5\pm3.6$  & $117.4\pm2.5$ \\
    2 layers& $27.4\pm9.1$ & $76.4\pm11.1$ & $119.0\pm10.5$ \\
    3 layers& $32.1\pm15.0$ & $78.5\pm15.0$ & $117.7\pm15.0$ \\
    \addlinespace[1.5ex]
    \multicolumn{4}{c}{\textbf{2D mask}} \\
    \addlinespace[0.8ex]
    1 layer & $24.3\pm3.6$  & $79.5\pm1.3$  & $120.0\pm1.7$ \\
    2 layers& $25.5\pm8.8$  & $75.9\pm11.7$ & $118.1\pm7.5$ \\
    3 layers& $34.1\pm15.5$ & $74.7\pm10$   & $118.0\pm18.6$ \\
    \bottomrule
  \end{tabular}
\end{table}
In the single-layer regions, where the model is expected to perform best, the three main orientations were retrieved with an error below about 4\textdegree\ for both masks. Consistently with the simulations, the model also performs reliably in the two layers case, providing average values close to the expected ones, although with an increase in the spread of the values. In contrast, for three layers the retrieval becomes less precise: the estimates of the average directions show larger deviations, particularly for the first orientation where errors of up to 7\textdegree\ are observed, and the overall uncertainty increases, reaching values as high as 18\textdegree. 

In conclusion, a DF model capable of retrieving multiple fibre orientations within a pixel from a radiographic DF image has been developed and validated with both simulated and experimental data. The model is based on a geometrical description, assuming known single-fibre scattering functions and equal scattering power for overlapping fibres. While this assumption may appear limiting, it is compatible with many real-world situations, such as carbon composite design, which is usually made of two stacked layers with equal thickness but different orientations. In addition, the directions of stacked fibres with different scattering power can also be retrieved if the corresponding $\sigma$ values are known a priori. These values can be obtained from mathematical modelling or measured from single-fibre regions if the composition of the sample is known. The results demonstrate robust orientation retrieval with good reliability for up to two fibre orientations per pixel but accuracy decreases for three orientations. Further improvements may be achievable through more advanced fitting approaches, for example, by incorporating a-priori knowledge of the sample structure using weighted or Bayesian curve fitting. Comparable results were obtained with both 1D and 2D masks. While the 1D mask requires sample rotation, the 2D mask provides multi-directional information in a single acquisition, offering a promising route toward time-resolved directional imaging with multi-orientation capability.  

\section*{Acknowledgment}
LM was supported by the EU Marie Sklodowska-Curie postdoctoral fellowship (grant agreement 101062305 BioDir-X). AO was supported by the Royal Academy of Engineering through their “Chairs in Emerging Technologies” scheme (CiET1819/2/78). This work was funded by Regione Puglia and CNR for Tecnopolo per la Medicina di Precisione D.G.R. n.2117 of 21/11/2018 Piano Nazionale di Ripresa e Resilienza (PNRR) - missione 4, componente
2, investimento 1.5, progetto ecs 00000024 Rome Technopole

\section*{Data availability}
The data that support the findings of this study are available from the corresponding author upon reasonable request.

\section*{Author contributions}
LM developed the model and designed the experiment and the simulations. LM, MF, SM, CR, AO performed the experiment. LM, MF, AO ,GG and AC analyzed the data. All authors discussed the results and contributed to writing the manuscript.

\bibliographystyle{unsrt}
\bibliography{Bibliografia.bib}

\end{document}